\documentclass{PoS}

\title{Precision Measurements of Fundamental Interactions 
with (Anti)Neutrinos
}

\ShortTitle{Precision (Anti)Neutrino Scattering off Nucleons and Nuclei}

\author{\speaker{R. Petti}\\
        Department of Physics and Astronomy, University of South Carolina, Columbia SC 29208, USA\\
        E-mail: \email{roberto.petti@cern.ch}}

\abstract{We discuss the main limitations of past neutrino scattering experiments and possible 
ways to address them in a next-generation program of precision measurements 
of fundamental interactions with (anti)neutrinos. A reduction of the longstanding precision 
gap with respect to electron scattering experiments could provide interesting synergies with 
the existing efforts in the fixed-target, collider, and nuclear physics communities.

          }

\FullConference{XXVII International Workshop on Deep-Inelastic Scattering and Related Subjects - DIS2019\\
		8-12 April, 2019\\
		Torino, Italy}

\begin{document}

\section{Limitations of Neutrino Scattering Experiments} 

Neutrinos and antineutrinos have many desirable properties including being 
sensitive only to weak interactions, full polarization, and a complete flavor separation 
($d/u, s/ \bar s, \bar d / \bar u$, valence/sea) through the Charged Current (CC) 
process. These features makes them an ideal probe for 
electroweak physics and for the structure of nucleons and nuclei. 
However, this unique potential has been only partially explored due to 
four major limitations of $\nu (\bar \nu)$ scattering experiments. 

\noindent 
\underline{\bf Statistics vs.\ resolution}: 
The need of massive detectors to compensate for the tiny cross-sections implies that 
neutrino experiments are usually a compromise between the collected statistics 
(target mass) and the resolutions achievable in the reconstruction of the interactions. 
The table below shows a summary of the main features of some of the 
past and present neutrino experiments. The largest samples correspond to 
about $10^7$ $\nu_\mu$ CC events. Only a few experiments achieved uncertainties 
on the muon energy scale, $\Delta E_\mu$, at the sub-percent level, at the price of 
smaller statistics. 

\noindent 
\underline{\bf Targets}: 
Most neutrino experiments are based upon massive detectors, consisting of 
several different materials and components, providing at the same time the 
active detection system and the neutrino target mass.  
A drawback of this configuration is the difficulty to precisely know the 
target chemical composition and mass, limiting the ultimate precision 
achievable in the measurements. Additional potential issues are the difficulty to 
modify the target configuration during data taking and/or the limited options 
for the target materials. 

\noindent 
\underline{\bf Fluxes}:
The energy of the incoming (anti)neutrino is unknown on an event-by-event basis and 
typically can vary over a broad range in conventional wide band neutrino beams. 
For this reason neutrino experiments have been affected by relatively large (5-10\%) 
systematic uncertainties on the knowledge of the (anti)neutrino fluxes. 

\noindent 
\underline{\bf Nuclear effects}:
Heavy nuclear targets are required to collect large statistics, but introduce 
additional systematic uncertainties with respect to elementary targets~\cite{Alvarez-Ruso:2017oui}. 
The unknown (anti)neutrino energy must be inferred from the detected final 
state particles originated in the interactions, which are affected by a substantial 
nuclear smearing. The corresponding unfolding process typically implies 
Monte Carlo and/or model corrections depending upon a number of different 
parameters, often empirically tuned with the observed 
kinematic distributions~\cite{Mosel:2019vhx}. 
These corrections can be critical in regions 
with small acceptance or large variations of the kinematics.

\begin{table}[b]
\begin{center} 
{\small \begin{tabular}{l|c|c|c|c|c|c}
Experiment & Mass & \multicolumn{1}{c|}{$\nu_{\mu}$ CC Stat.} & Target & $E_\nu$ (GeV)
& $\Delta E_\mu$  & $\Delta E_{\rm H}$ \\ \hline\hline
            CDHS  &  750 t &  { $10^{7}$}   &  p,Fe & 20-200 & 2.0\% & 2.5\% \\
            BEBC  & various &   5.7$\times$$10^{4}$   & p,D,Ne & 10-200 &  & \\
            CCFR  & 690 t & { 1.0$\times$$10^{6}$}   & Fe & 30-360 & 1.0\% & 1.0\% \\
            NuTeV  &  690 t & 1.3$\times$$10^{6}$  & Fe & 30-360 & 0.7\% & 0.43\% \\
            CHORUS  & 100 t & 3.6$\times$$10^{6}$   & Emul.,Pb & 10-200 & 2.5\% & 5.0\% \\
            NOMAD  & 2.7 t & 1.5$\times$$10^{6}$   & C,Fe & 5-200 & 0.2\% & 0.5\% \\
            MINOS ND & 980 t &  3.6$\times$$10^{6}$   &  Fe & 3-50 & 2-4\% & 5.6\% \\
            T2K ND &  1.9 t & $10^5$  & CH,H$_2$O  &  0.2-5 & 0.6\%  &  2-4\% \\
            MINER$\nu$A &  5.4 t  & 10$^7$  & CH,C,Fe,Pb  & 1-30  & 2\% &   \\
            \hline
\end{tabular}} 
\vspace*{-1.00cm}
\end{center} 
\end{table} 

As a result, a major precision gap is present between (anti)neutrino experiments and 
the electron experiments performed at JLab, HERA, LEP, SLAC, etc. 
The four issues described above represent intrinsic limitations 
of current (anti)neutrino scattering experiments, making an accurate detection of (anti)neutrino 
interactions extremely challenging. This difficulty is illustrated by the 
many outstanding discrepancies among existing 
measurements performed by different experiments, 
as well as with various theoretical models~\cite{Alvarez-Ruso:2017oui}.  
Any sensible program of precision measurements of fundamental interactions with 
(anti)neutrinos must first address each of those four bottlenecks.

\vspace*{-0.20cm}
\section{New Perspectives on Precision (Anti)Neutrino Scattering} 

\vspace*{-0.20cm}
\subsection{Statistics and Resolution} 

\begin{figure}[t]
\centering
\vspace*{-0.50cm}
\includegraphics[width=1.00\textwidth]{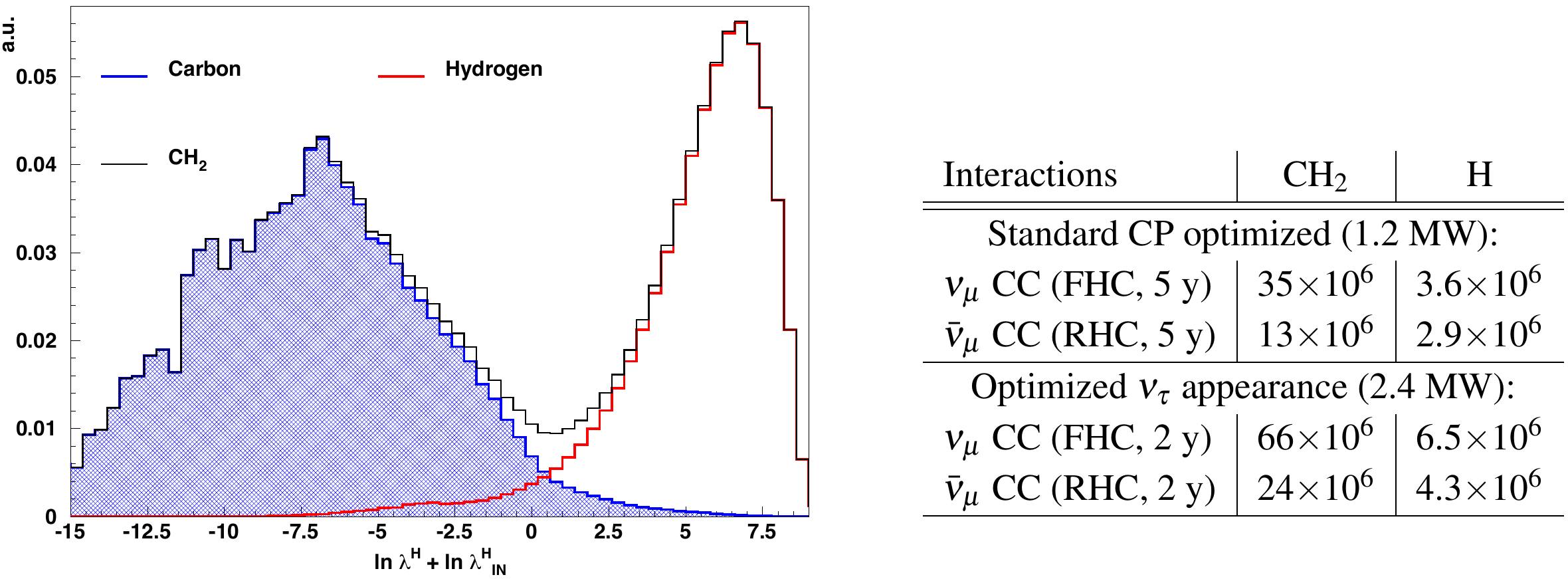} 
% \hspace*{0.50cm} \begin{tabular}{l|c|c}
% Interactions & CH$_2$  & H  \\ \hline\hline
% \multicolumn{3}{c}{Standard CP optimized (1.2 MW):} \\
% $\nu_\mu$ CC (FHC, 5 y) & 35$\times$10$^6$  & 3.6$\times$10$^6$ \\
% $\bar \nu_\mu$ CC (RHC, 5 y) & 13$\times$10$^6$ & 2.9$\times$10$^6$  \\ 
% \hline
%\multicolumn{3}{c}{Optimized $\nu_\tau$ appearance (2.4 MW):} \\
% $\nu_\mu$ CC (FHC, 2 y) & 66$\times$10$^6$  &  6.5$\times$10$^6$ \\
% $\bar \nu_\mu$ CC (RHC, 2 y) & 24$\times$10$^6$ & 4.3$\times$10$^6$ \\ 
% \hline 
% \end{tabular}
\vspace*{-0.80cm}
\caption{%
Left panel: Example of kinematic identification of $\nu_\mu p \to \mu^- p \pi^+$ 
interactions on hydrogen using multi-dimensional likelihood functions in STT~\cite{Duyang:2018lpe}. 
Right panel: Expected CC statistics in STT (5 tons of CH$_2$) with the various LBNF beam options considered~\cite{ESGprop}. 
\label{fig:STT-Hsel}}
\end{figure}

The Long-Baseline Neutrino Facility (LBNF) is expected to provide $\nu \& \bar \nu$ 
beams of unprecedented intensity, with a nominal beam power of 
1.2 MW and $1.1 \times 10^{21}$ pot/year. Although primarily designed for the 
study of long-baseline neutrino oscillations, this facility offers a unique opportunity 
for neutrino scattering physics~\cite{Petti2018,ESGprop}, due to the availability of 
a high-energy beam option optimized to detect the $\nu_\tau$ appearance from neutrino 
oscillations, in addition to the default low-energy beam optimized for the search for CP violation. 
It is conceivable to have a dedicated 2 year run with this high energy 
beam after the completion of the nominal data taking of 5+5 years with the default 
low-energy neutrino and antineutrino beams~\cite{Petti2018,ESGprop}. 
By that time the planned upgrades of the beam intensity to a nominal power 
of 2.4 MW would more than double the available statistics. Under such conditions, 
a detector with a fiducial mass of about 5 tons would collect more than 
$10^8$ CC interactions at the near site (Fig.~\ref{fig:STT-Hsel}). Given this relatively 
compact size, we can afford a high resolution detector providing an accurate measurement 
of the four-momenta of all final state particles produced in (anti)neutrino interactions and 
$\Delta E_\mu \leq 0.2\%$. 

\begin{figure}[t]
\centering
\vspace*{-0.50cm}
\includegraphics[width=0.80\textwidth]{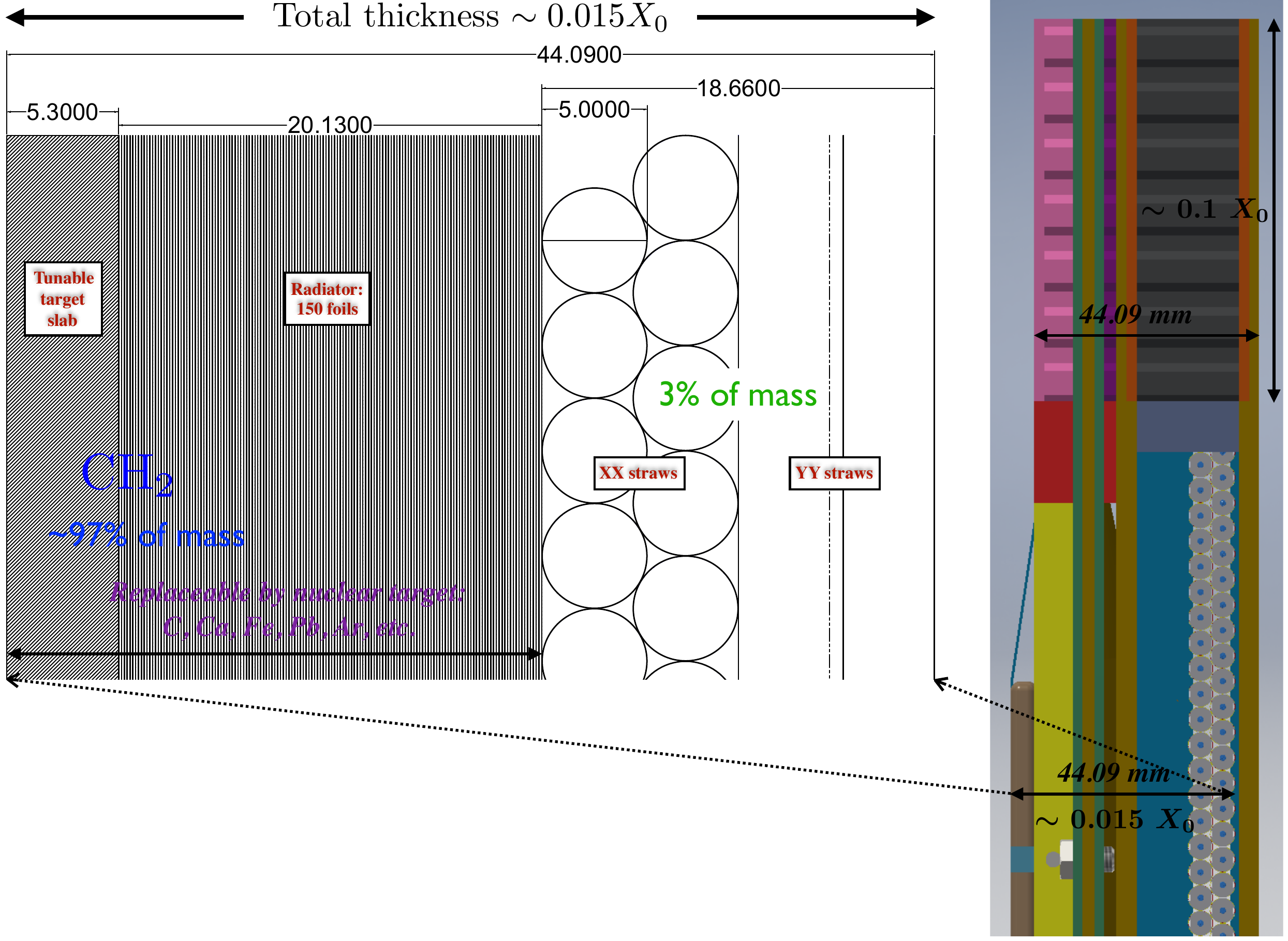}
\vspace*{-0.30cm}
\caption{%
Drawing of a STT module allowing a control of the configuration, chemical composition, and mass of the $\nu(\bar \nu)$ target(s) comparable to electron scattering experiments. 
\label{fig:CompactSTT}}
\end{figure}

\vspace*{-0.20cm}
\subsection{Control of Targets} 

A detector technology designed to offer a control of the configuration, chemical 
composition and mass of the neutrino targets similar to the one achieved in 
electron scattering experiments is a Straw Tube Tracker (STT), 
in which the neutrino targets are physically 
separated from the actual tracking system of negligible mass~\cite{Petti2004}. 
To achieve high resolution measurements, 
the target mass is spread out uniformly throughout the entire tracking volume, 
by keeping the  average density low enough 
-- similar to that of liquid deuterium -- 
to have a detector transparent (size comparable to one radiation length)  
to final state particles produced in neutrino interactions. 
This detector concept requires to be inserted into a magnetic field and 
surrounded by a $4\pi$ electromagnetic calorimeter (ECAL). 
Examples of different implementations of the STT technology can be found in the 
CDR of the DUNE experiment~\cite{Mishra:2008nx,Acciarri:2016ooe} and in a recently proposed 
near detector enhancement for the Long-Baseline Neutrino Facility (LBNF)~\cite{ESGprop}. 

Figure~\ref{fig:CompactSTT} shows a new STT design offering a compact detector 
equipped with fully configurable neutrino targets,  as well as 
optimized tracking and particle identification.
The base tracking technology is provided by 
low-mass straws similar to the ones used in many modern experiments 
for precision physics or the search for rare 
processes~\cite{Sergi:2012twa,Anelli:2015pba,Nishiguchi:2017gei,Lee20162530,Gianotti:2013ipa}. 
Thin layers -- typically 1-2\% of radiation length -- of various target 
materials (100\% chemical purity) are alternated with straw layers so that they 
represent about 97\% of the total detector mass 
(the mass of the straws being $3\%$). This feature, combined 
with the excellent vertex, angular, momentum, and timing 
resolutions are key factors to correctly associate 
neutrino interactions to each target material, as well as for an 
accurate measurement of the four-momenta of the final state particles. 
The main target material is CH$_2$ plastic (polypropylene) split between a 
tunable solid target slab and a radiator (150 foils 15 $\mu m$ thick) for electron 
identification via transition radiation (Fig.~\ref{fig:CompactSTT}). Both the target slab and 
the radiator can be easily unmounted (or replaced) during data taking, leaving a pure tracking module 
with average density of $\sim$0.005 g/cm$^3$. The average density (and fiducial mass) 
of the detector can be fine tuned up to a maximal value of about $\sim$0.18 g/cm$^3$. 
The system composed of the target slab and the radiator can also be replaced by a 
variety of thin ($\ll 0.1X_0$) nuclear targets like C, Ca, Fe, Pb, Ar, etc., which are interleaved 
with CH$_2$ modules to guarantee the same acceptance. 

\vspace*{-0.20cm}
\subsection{Control of Nuclear Effects} 

The accurate control of the configuration, chemical composition and mass of the 
(anti)neutrino targets provided by the STT allows the implementation of a solid 
hydrogen target by subtracting measurements on dedicated graphite (pure C) 
targets from those on the CH$_2$ plastic targets described above~\cite{Duyang:2018lpe}. 
The high resolution of STT allows the identification of the   
interactions on hydrogen within the CH$_2$ target before subtraction by using a kinematic analysis. 
Since the H target is at rest, the Charged Current (CC) events are expected to be 
perfectly balanced in a plane transverse to the beam direction (up to the tiny beam divergence) 
and the muon and hadron vectors are back-to-back in the same plane. Instead, events from 
nuclear targets are affected by both initial and final state nuclear effects, resulting in 
a significant missing transverse momentum and a smearing of the transverse plane kinematics. 
By exploit these differences using the reconstructed event kinematics we can achieve 
efficiencies exceeding 90\% and purities of 80-95\% (Fig.~\ref{fig:STT-Hsel})~\cite{Duyang:2018lpe}.  
The subtraction of the small residual background is entirely data-driven by using the 
corresponding graphite target measurements, which automatically 
include all types of interactions, as well as reconstruction effects, relevant for the 
hydrogen selection. 

The availability of a solid hydrogen target provides 
high statistics $\nu(\bar \nu)$-H CC samples (Fig.~\ref{fig:STT-Hsel}) 
free from nuclear effects, to be compared with the corresponding ones obtained from 
the nuclear targets installed within the same detector (same acceptance). 
This comparison can be performed with both inclusive CC events and with 
exclusive topologies and allows a direct measurement of nuclear effects 
from initial and final state interactions, constraining the corresponding 
nuclear smearing on the detected interactions. As a result, we can reduce the 
uncertainties in the unfolding of data collected from heavy targets and 
precisely calibrate the reconstructed neutrino energy scale.

\vspace*{-0.20cm}
\subsection{Control of Fluxes} 

\begin{figure}[t]
\centering
\vspace*{-0.50cm}
\includegraphics[width=1.00\textwidth]{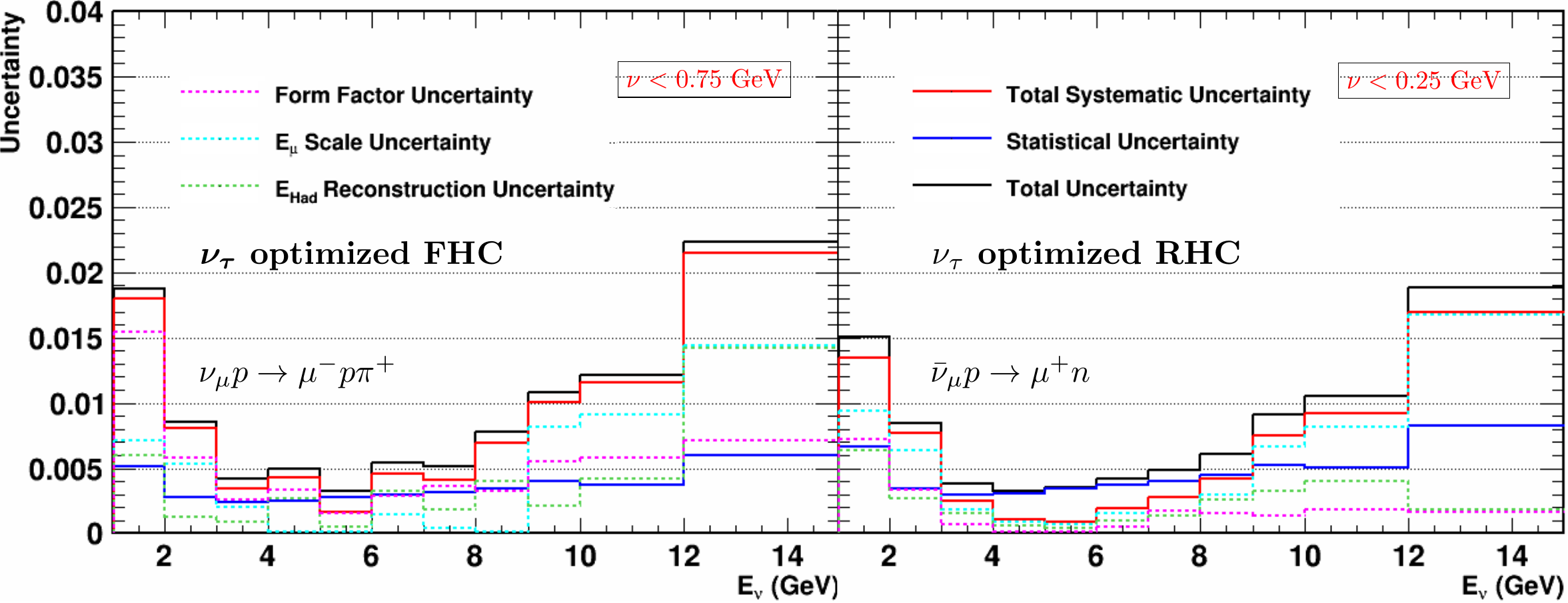} 
\vspace*{-0.80cm}
\caption{%
Accuracy in the determination of the $\nu_\mu$ (left panel) and $\bar \nu_\mu$ (right panel) 
relative fluxes using $\nu_\mu p \to \mu^- p \pi^+$ and $\bar \nu_\mu p \to \mu^+ n$ exclusive 
processes on H with small energy transfer $\nu$ in STT~\cite{Duyang:2019prb}. 
\label{fig:STT-Hflux}}
\end{figure}

The solid hydrogen target allows the determination of $\nu_\mu$ and $\bar \nu_\mu$ relative 
fluxes with an accuracy better than 1\%  
in conventional wide-band neutrino beams, using exclusive $\nu_\mu p \to \mu^- p \pi^+$, 
$\bar \nu_\mu p \to \mu^+ p \pi^-$, and $\bar \nu_\mu p \to \mu^+ n$ processes on 
hydrogen with small energy transfer $\nu$~\cite{Duyang:2019prb}. 
This level of precision cannot be achieved with other known techniques using 
nuclear targets. The hydrogen target solves the problems arising from the 
nuclear smearing in conventional targets, while the small energy transfer 
reduces the systematic uncertainties on the energy dependence of the cross-sections. 
All relevant systematic uncertainties affecting the flux measurements can be directly 
constrained using data themselves. Figure~\ref{fig:STT-Hflux} shows the 
achievable precisions on the $\nu_\mu$ and $\bar \nu_\mu$ relative fluxes 
with the high-energy LBNF beam option. 
The measurement of the $\bar \nu_\mu p \to \mu^+ n$ interactions on H at small 
momentum transfer $Q$ also provides the absolute $\bar \nu_\mu$ flux, 
since the corresponding cross-section in the limit $Q\to 0$ is a constant known 
to high accuracy from neutron $\beta$ decay~\cite{Duyang:2019prb}. 
The absolute $\nu_\mu$ flux can be accurately determined with the 
$\nu e^- \to \nu e^-$ elastic scattering, by exploiting the excellent electron identification 
and angular resolution of STT.

\vspace*{-0.20cm}
\section{Precision Measurements and Searches} 

We discussed a possible way to address the main limitations of neutrino scattering 
experiments -- statistics and resolution, control of targets, nuclear effects, and fluxes -- 
largely reducing the precision gap with respect to electron scattering experiments. 
These improvements would allow us to exploit the unique properties of the (anti)neutrino 
probe for precision studies of fundamental interactions and of the structure of 
nucleons and nuclei~\cite{Petti2018,ESGprop}. 
The near site of the LBNF could then be turned into a general 
purpose $\nu \& \bar \nu$ physics facility with a broad program of physics 
measurements complementary to the ongoing efforts in the fixed-target~\cite{Dudek:2012vr}, 
collider~\cite{Accardi:2012qut}, and nuclear physics communities. 
The level of precision enabled by the techniques described above can provide 
insights on various fields, unveiling some discovery potential and 
generating hundreds of diverse physics studies. 
The same features can also help to reduce the systematic uncertainties affecting the 
long-baseline neutrino oscillations measurements. 

Precision tests of electroweak physics can be performed using complementary 
channels including the ratio of neutral to charged current (NC/CC) interactions 
in $\nu$-N DIS, $\nu e^-$ NC elastic scattering, ratios of NC elastic scattering 
off protons to CC quasi-elastic scattering, NC/CC ratio of coherent $\rho$ production. 
The unique combination of hydrogen and nuclear targets allows precision studies of 
the structure of nucleons and nuclei exploiting the flavor selection of the weak current: 
structure functions and parton distributions, QCD studies, sum rules, 
high twists and non-perturbative effects, strangeness content of the nucleon, 
charm production, isospin physics, structure of the weak current, etc. 
Examples are given by precision tests of the Adler~\cite{Adler:1964yx} and 
Gross-Llewellyn Smith~\cite{Gross:1969jf} sum rules on hydrogen and in 
nuclei~\cite{Kulagin:2007ju,Kulagin:2004ie}. 
Similarly, by exploiting the isospin symmetry $F_2^{\nu n} = F_2^{\bar \nu p}$ 
on hydrogen, we can obtain a model-independent measurement of the free neutron 
structure functions, as well as a measurement of the large $x$ behavior of the $d/u$ quark
ratio~\cite{Alekhin:2017fpf}. These measurements can also be used for 
precision tests of the isospin (charge) symmetry and would help to elucidate the 
flavor structure of the nucleon~\cite{Alekhin:2018dbs}. 
The extensive program of precision measurements described above would concurrently 
be sensitive to new physics beyond the Standard Model, in a way complementary to direct searches. 
These latter can also be performed over a broad range of topics like the 
MiniBooNE anomaly, $\nu_\tau$ appearance and non-standard interactions, Dark Sector physics 
like heavy sterile neutrinos (e.g. Majorana singlet fermions in $\nu$MSM models), 
axion-like particles, dark photons, light (sub-GeV) dark matter, etc.

\end{document}